\documentclass[aps,prb,preprint,superscriptaddress]{revtex4-2}

\usepackage[utf8]{inputenc} 
\usepackage{graphicx}
\usepackage{amsmath}
\usepackage{bm}
\usepackage{dcolumn}
\usepackage{multirow}
\usepackage{booktabs}

\begin{document}


\title{Phonon-limited carrier mobilities and Hall factors in 4H-SiC from first principles}


\author{Tianqi Deng}
\email{dengtq@zju.edu.cn}
\author{Deren Yang}
\author{Xiaodong Pi}
\email{xdpi@zju.edu.cn}
\affiliation{State Key Laboratory of Silicon Materials and School of Materials Science and Engineering, Zhejiang University, Hangzhou, Zhejiang 310027, China}
\affiliation{Institute of Advanced Semiconductors \& Zhejiang Provincial Key Laboratory of Power Semiconductor Materials and Devices, ZJU-Hangzhou Global Scientific and Technological 
Innovation Center, Hangzhou, Zhejiang 311200, China}

\date{\today}

\begin{abstract}
Charge carrier mobility is at the core of semiconductor materials and devices optimization, and Hall measurement is one of the most important techniques for its characterization. The Hall factor, defined as the ratio between Hall and drift mobilities, is of particular importance. Here we study the effect of anisotropy by computing the drift and Hall mobility tensors of a technologically important wide-band-gap semiconductor, 4H-silicon carbide (4H-SiC) from first principles. With $GW$ electronic structure and \textit{ab initio} electron-phonon interactions, we solve the Boltzmann transport equation without fitting parameters. The calculated electron and hole mobilities agree with experimental data. The electron Hall factor strongly depends on the direction of external magnetic field $\mathbf{B}$, and the hole Hall factor exhibits different temperature dependency for $\mathbf{B}\parallel c$ and $\mathbf{B}\perp c$. We explain this by the different equienergy surface shape arising from the anisotropic and non-parabolic band structure, together with the energy-dependent electron-phonon scattering.
\end{abstract}


\maketitle

\section{Introduction\label{sec1}}

Increasing demands in high-power and high-temperature electronic devices call for wide-band-gap semiconductors as alternative functional materials to silicon. Silicon carbide (SiC) has become one of the most promising materials in power electronic devices owning to its unique combination of high carrier mobility, high critical field strength, high saturation velocity, and high thermal conductivity \cite{Kimoto2014, Millan2014, Kimoto2015, Puschkarsky2019, Han2021, Tian2021}. Among the more than two hundred polytypes, 4H-SiC is preferred for its wider band gap and higher critical electric field than the cubic 3C-SiC and higher carrier mobilities and lower anisotropy as compared to 6H-SiC. Therefore, it is more technologically relevant and has become the major functional SiC polytype for applications in electronic devices \cite{Kimoto2014}.

Despite the recent surge of academic and industrial interests in 4H-SiC, many important aspects of its physical properties and the underlying physics are not clarified yet. For example, as a hexagonal crystal, anisotropy is expected for its physical properties like mechanical \cite{Liu2022} and transport properties \cite{Kimoto2014} such as carrier mobilities and Hall effect. The carrier mobility is a key functional property that determines device performance such as on-resistance \cite{Huang2022}. However, it is difficult to experimentally distinguish between the anisotropy contributions from drift mobility and Hall factor in a Hall measurement, and the common practice is either assuming a unity Hall factor $r_H=1$, estimating $r_H$ using empirically parametrized models, or estimating the true carrier concentration from dopant concentration and activation energies \cite{Iwata2001,Parisini2013,Asada2016,Tanaka2018,Ishikawa2021, Huang2022}. Additionally, the analysis of one of most important mechanisms underlying its charge transport phenomena, i.e. the electron-phonon interactions and scatterings, still relies on empirically determined, adjustable parameters with significant uncertainty. These adjustable parameters were also employed to explain exotic phenomena such as non-unity hole Hall factors \cite{Iwata2001,Asada2016,Tanaka2018}. Therefore, it becomes increasingly important to investigate such microscopic physics and confirm their respective contributions in the charge transport process without resorting to uncertain fitting parameters.

Electron-phonon interactions from the density functional perturbation theory (DFPT) calculations \cite{Giustino2017} emerged as a powerful tool for studying importance phenomena in solid state and their underlying microscopic mechanisms, including phonon-limited charge transport \cite{Ponce2019,Ponce2020a,Ponce2021}, superconductivity \cite{Ma2021,Giustino2007a}, polaron \cite{Verdi2017,Sio2019a}, phonon-assisted optical absorption \cite{Noffsinger2012}, band structure renormalization \cite{Giustino2010,Li2020}, etc. In conjunction with Boltzmann transport equation (BTE), the charge transport in the presence of electrical field and magnetic field can be simulated self-consistently to obtain key quantities like drift mobility \cite{Ponce2019}, breakdown field \cite{Ponce2020}, and thermoelectricity \cite{Deng2020}. Recently, the Hall effect in several typical cubic semiconductors has been studied by solving the BTE in the presence of both electric and magnetic fields, where quantitative agreement has been achieved in comparison with experimental measurements \cite{Ponce2021}. It is thus intriguing to explore the possible anisotropy in the hexagonal phase, to compute the Hall factors in the intrinsic limit, and to clarify the role of electron-phonon interactions in 4H-SiC.

In this work, we performed in-depth analysis of electron-phonon interactions, phonon-limited charge transport, and their anisotropy in 4H-SiC by first-principles calculations. Both short-ranged and long-ranged dipolar/quadrupolar electron-phonon interactions are included from first principles in combination with Wannier-interpolation technique \cite{Brunin2020,Brunin2020a,Park2020,Jhalani2020,Ponce2021}. We find that the spin-orbit coupling (SOC) significantly affect the hole effective masses, even though the SOC splitting is small. The phonon-limited mobilities agree well with experimental measurements of lightly-doped samples, and hole mobility exhibits a stronger anisotropy than that of electron. The electrons are mainly scattered by optical phonons, while the band-edge holes are mostly scattered by acoustic phonons. The Hall factors depend on the directions of both the applied magnetic field and the electric current. Hall factors deviate from 1 for both electrons and holes, and distinct temperature-dependence were predicted. The non-unity is explained by non-parabolic band structure, non-spherical equienergy surface, and energy-dependent electron-phonon scattering strength. This work thus clarifies the anisotropic charge transport phenomena in 4H-SiC and the impact of electron-phonon interactions in the intrinsic limit from a microscopic, \textit{ab initio} perspective. The predicted Hall factors without empirical, adjustable parameters also allows possible comparison between drift mobility and Hall mobility from experimental measurements.

\section{Methods\label{sec2}}

\begin{figure}
\includegraphics[width=8.6cm]{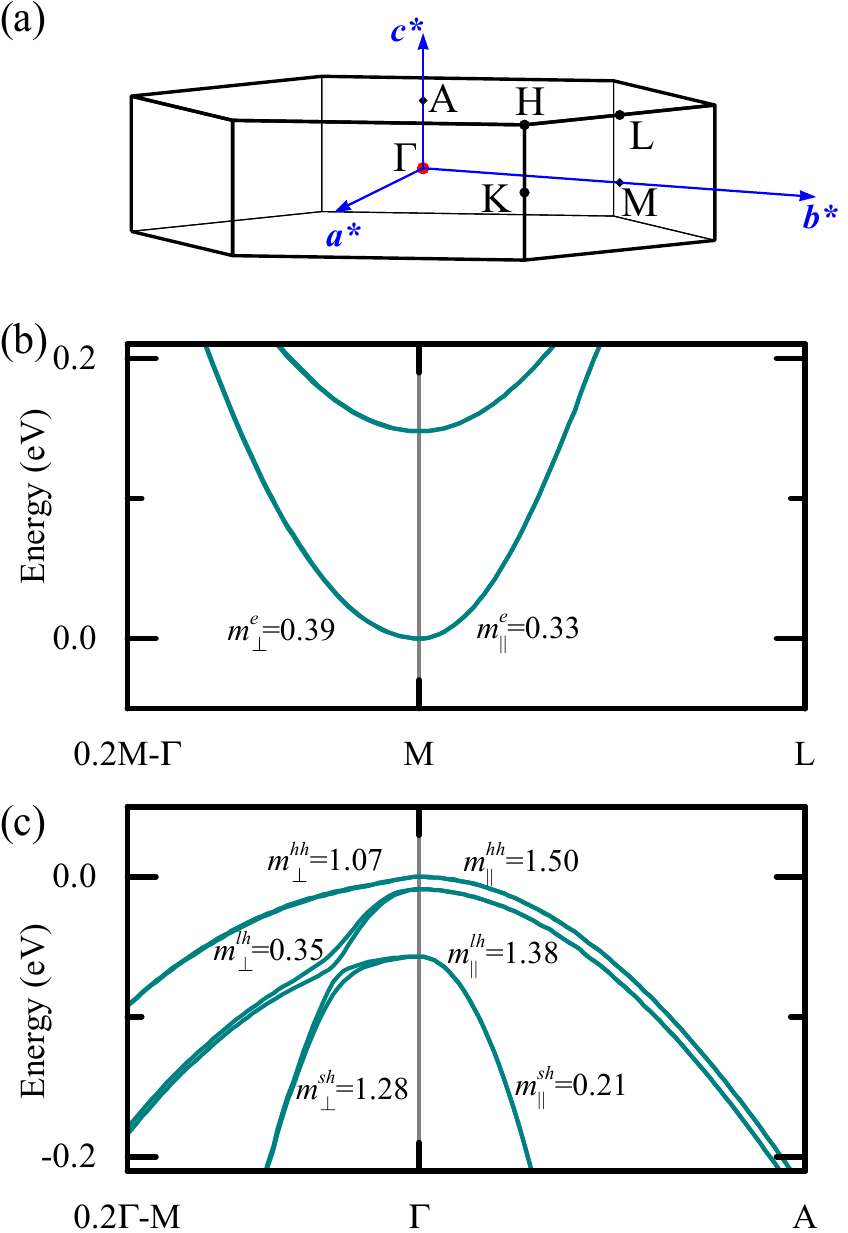}
\caption{\label{bzband} (a) The first Brillouin zone of 4H-SiC and major high-symmetry points. The band structure of (b) conduction and (c) valence bands near the band edges. The effective masses along ($\parallel$) and perpendicular to the $c$-axis ($\perp$) are also given.}
\end{figure}

\subsection{Carrier mobility and Hall effect calculations}
In a typical Hall measurement for Hall mobility along $\alpha$ direction, an electric current density $\mathbf{j}$ along $\alpha$ and magnetic field $\mathbf{B}$ along $\gamma$ are applied, and the induced Hall field $\mathbf{E}$ or Hall voltage along $\beta$ is measured, where $\alpha$, $\beta$ and $\gamma$ are orthogonal. In the linear regime of small $B_{\gamma}$, the Hall coefficient is
\begin{align}
R_{\alpha\beta\gamma}^{H}=\frac{E_{\beta}}{j_{\alpha}B_{\gamma}} = \frac{[\sigma^{-1}(B_{\gamma})-\sigma^{-1}(0)]_{\beta\alpha}}{B_{\gamma}}\nonumber\\
\approx\left[\sigma^{-1}(0)\frac{\sigma(B_{\gamma})-\sigma(0)}{B_{\gamma}}\sigma^{-1}(0)\right]_{\beta\alpha}\label{eq:rh}.
\end{align}
Therefore, calculation of the carrier mobility and Hall coefficient involves computing the magnetic field $\mathbf{B}$-dependent conductivity tensor
\begin{equation}
[\sigma(\mathbf{B})]_{\alpha\beta}=\frac{-e}{V_{uc}}\sum_{n}\int_{\text{BZ}}\frac{d^3 k}{\Omega_{\text{BZ}}}v_{n\mathbf{k}\alpha}\partial_{E_{\beta}}f_{n\mathbf{k}}^{\mathbf{B}}.
\end{equation}

Here $V_{uc}$ is the unit-cell volume, $v_{n\mathbf{k}\alpha}=\frac{\partial \varepsilon_{n\mathbf{k}}}{\partial \hbar k_{\alpha}}$ is the band velocity defined as the $\mathbf{k}$-derivative of eigen-energy $\varepsilon_{n\mathbf{k}}$ along $\alpha$ direction, and $\partial_{E_{\beta}}f_{n\mathbf{k}}^{\mathbf{B}}$ is the solution of the linearized (BTE) with magnetic field $\mathbf{B}$
\begin{align}
&-ev_{n\mathbf{k}\beta}\frac{\partial f_{n\mathbf{k}}^0}{\partial\varepsilon_{n\mathbf{k}}}-\frac{e}{\hbar}(\mathbf{v}_{n\mathbf{k}}\times \mathbf{B})\cdot\mathbf{\nabla_{\mathbf{k}}}\partial_{E_{\beta}}f_{n\mathbf{k}}^{\mathbf{B}}\nonumber\\
=&\sum_{m}\int_{\text{BZ}}\frac{d^3 q}{\Omega_{\text{BZ}}}\left[\tau_{m\mathbf{k+q}\to n\mathbf{k}}^{-1}\partial_{E_{\beta}}f_{m\mathbf{k+q}}^{\mathbf{B}}\right.\nonumber\\
&\left.-\tau_{n\mathbf{k}\to m\mathbf{k+q}}^{-1}\partial_{E_{\beta}}f_{n\mathbf{k}}^{\mathbf{B}}\right],
\end{align}
with $f_{n\mathbf{k}}^0$ and $\Omega_{\text{BZ}}$ being the equilibrium Fermi-Dirac distribution and first Brillouin zone volume, and electron-phonon (\textit{e-ph}) scattering rate defined as
\begin{align}
&\tau_{n\mathbf{k}\to m\mathbf{k+q}}^{-1}\nonumber\\
=&\frac{2\pi}{\hbar}\sum_{\nu}|g_{mn\nu}(\mathbf{k},\mathbf{q})|^2\nonumber\\
&\times[(n_{\nu\mathbf{q}}+1-f_{m\mathbf{k+q}}^0)\delta(\varepsilon_{n\mathbf{k}}-\varepsilon_{m\mathbf{k+q}}-\hbar\omega_{\nu\mathbf{q}})\nonumber\\
&+(n_{\nu\mathbf{q}}+f_{m\mathbf{k+q}}^0)\delta(\varepsilon_{n\mathbf{k}}-\varepsilon_{m\mathbf{k+q}}+\hbar\omega_{\nu\mathbf{q}})]\label{eq:tau}.
\end{align}
The \textit{e-ph} matrix element $g_{mn\nu}(\mathbf{k},\mathbf{q})$ is induced by a phonon $\nu\mathbf{q}$ that scatters an electron from state $\left|n\mathbf{k}\right>$ to $\left|m\mathbf{k+q}\right>$, and $n_{\nu\mathbf{q}}$ is the equilibrium Bose-Einstein distribution. The BTE can be further re-cast into a form that is more suitable for iterative solution
\begin{align}
&\left [1-\frac{e}{\hbar}\tau_{n\mathbf{k}}(\mathbf{v}_{n\mathbf{k}}\times \mathbf{B})\cdot\mathbf{\nabla_{\mathbf{k}}}\right ]\partial_{E_{\beta}}f_{n\mathbf{k}}^{\mathbf{B}}\nonumber\\
=&ev_{n\mathbf{k}\beta}\tau_{n\mathbf{k}}\frac{\partial f_{n\mathbf{k}}^0}{\partial\varepsilon_{n\mathbf{k}}}\nonumber\\
&+\frac{2\pi\tau_{n\mathbf{k}}}{\hbar}\sum_{m}\int_{\text{BZ}}\frac{d^3 q}{\Omega_{\text{BZ}}}\left[\tau_{m\mathbf{k+q}\to n\mathbf{k}}^{-1}\partial_{E_{\beta}}f_{m\mathbf{k+q}}^{\mathbf{B}}\right]\label{eq:bte},
\end{align}
The total scattering time is
\begin{equation}
\tau_{n\mathbf{k}}^{-1}=\sum_{m}\int_{\text{BZ}}\frac{d^3 q}{\Omega_{\text{BZ}}}\tau_{n\mathbf{k}\to m\mathbf{k+q}}^{-1}.
\end{equation}
The above equations can be solved from first principles without any empirical parameters \cite{Ponce2020a,Ponce2021,Desai2021,Macheda2018}. By solving Eq. \eqref{eq:bte} iteratively one obtain the self-consistent solution of $\partial_{E_{\beta}}f_{n\mathbf{k}}^{\mathbf{B}}$. One common approximate solution is the so-called self-energy relaxation time approximation (SERTA), which is approximated using the scattering time $\tau_{n\mathbf{k}}$ by omitting the second term on the right-hand side of Eq. \eqref{eq:bte}. This is equivalent to the first step in iterative solution of Eq. \eqref{eq:bte}. Further approximation, i.e. the constant relaxation time approximation (CRTA) can be made by assuming $\tau_{n\mathbf{k}}$ to be a universal constant $\tau_0$ such that only band structure information is included.

Given the Hall coefficient and conductivity tensor, the Hall mobility along $\alpha$ direction with $B_{\gamma}$ can be computed as
\begin{equation}
\mu_{\alpha}^{\text{H},B_\gamma}=\sigma_{\alpha\alpha}R_{\alpha\beta\gamma}\label{eq:muh},
\end{equation}
and the drift mobility is
\begin{equation}
\mu_{\alpha}=\frac{\sigma_{\alpha\alpha}}{ne}\label{eq:mu}.
\end{equation}
Thus the Hall factor is the factor between Hall and drift mobility
\begin{equation}
r_{\alpha\beta\gamma}^{\text{H}}=\frac{\mu_{\alpha}^{\text{H},B_\gamma}}{\mu_{\alpha}}=R_{\alpha\beta\gamma}ne
\end{equation}

\subsection{First principles calculations}

Evaluation of the \textit{e-ph} scattering rate in Eq. \eqref{eq:tau} requires knowledge of $g_{mn\nu}(\mathbf{k},\mathbf{q})$ and $\varepsilon_{n\mathbf{k}}$ in the whole Brillouin zone. Here we compute the band structure $\varepsilon_{n\mathbf{k}}$ using density functional theory on a $8\times 8\times 2$ $\mathbf{k}$-grid, which is sufficient for accurate Wannier-interpolation as shown in the Appendix. The $g_{mn\nu}(\mathbf{k},\mathbf{q})$ on the same $\mathbf{k/q}$-grid were calculated using relativistic DFPT \cite{Giustino2017} as implemented in {\sc Quantum ESPRESSO} \cite{Giannozzi2009,Giannozzi2017}. Subsequent Fourier-Wannier-interpolation onto much finer grids \cite{Giustino2007} using maximally localized Wannier functions \cite{Marzari2012,Pizzi2020} were carried out with EPW software \cite{Noffsinger2010,Ponce2016}. Dipolar and quadrupolar corrections to the interpolation were also included \cite{Verdi2015,Ponce2021,Brunin2020,Brunin2020a,Jhalani2020,Park2020}, with Born effective charge and dielectric, and dynamical quadrupole tensors calculated from linear response \cite{Baroni2001,Royo2019} as implemented in {\sc Quantum ESPRESSO} \cite{Giannozzi2009,Giannozzi2017} and {\sc abinit} \cite{Gonze2020,Romero2020}. PBEsol generalized gradient approximation \cite{Perdew2008} and norm-conserving pseudopotentials from PseudoDojo project \cite{VanSetten2018} were used for structural relaxation and DFPT calculations. The $GW$ quasiparticle band structure $\varepsilon_{n\mathbf{k}}$ is computed using YAMBO \cite{Marini2009,Sangalli2019} with 800 bands for screening and Green's function calculations. Hybrid functional band structure was computed with HSE06 functional \cite{Heyd2003,Heyd2006}.

\section{Results and Discussion\label{sec3}}

\subsection{Band Structure}

\begin{table}
\caption{\label{table1} The effective masses of electrons ($e$) and holes ($hh$ for heavy hole, $lh$ for light hole, and $sh$ for crystal-field split-off hole) along transverse ($\perp$, in basal plane) and longitudinal ($\parallel$, along $c$-axis) directions. All masses are given in the unit of electron rest mass $m_e$.}
\begin{ruledtabular}
\begin{tabular}{ccccccc}
 \multicolumn{2}{c}{} & \multicolumn{2}{c}{w/o SOC} & \multicolumn{2}{c}{w/ SOC} & \\
 & & \multicolumn{1}{c}{{HSE}} & \multicolumn{1}{c}{{GW}} & \multicolumn{1}{c}{{HSE}} & \multicolumn{1}{c}{{GW}} & \multicolumn{1}{c}{{Exp}} \\ \midrule
 & $m_{\text{M}\Gamma}^{e}$ & 0.58 & 0.54 & 0.58 & 0.54 & 0.58±0.01 \cite{Volm1996}\\ \cmidrule{2-7}
{$m_{\perp}^{e}$} & $m_{\text{MK}}^{e}$ & 0.28 & 0.29 & 0.28 & 0.29 & 0.31±0.01 \cite{Volm1996} \\ \cmidrule{2-7}
 & $m_{\perp}^{e}$ & 0.4 & 0.39 & 0.4 & 0.39 & 0.425 \cite{Volm1996}\\ \cmidrule{1-7}
{$m_{\parallel}^{e}$} & $m_{\parallel}^{e}$ & 0.34 & 0.33 & 0.34 & 0.33 & 0.33±0.01 \cite{Volm1996} \\ \cmidrule{1-7}
 & $m_{\perp}^{hh}$ & 3.04 & 12.9 & 0.64 & 0.65 & {0.66±0.02 \cite{Son2000}} \\ \cmidrule{2-6}
 {$m_{\perp}^{h}$} & $m_{\perp}^{lh}$ & 0.29 & 0.31 & 0.45 & 0.44 & \\ \cmidrule{2-6}
 & $m_{\perp}^{sh}$ & 1.37 & 1.14 & 1.40 & 1.42 & \\ \cmidrule{1-7}
 & $m_{\parallel}^{hh}$ & 1.48 & 1.48 & 1.48 & 1.50 & {1.75±0.02 \cite{Son2000}} \\ \cmidrule{2-6}
 {$m_{\parallel}^{h}$} & $m_{\parallel}^{lh}$ & 1.48 & 1.48 & 1.30 & 1.35 & \\ \cmidrule{2-6}
 & $m_{\parallel}^{sh}$ & 0.20 & 0.20 & 0.20 & 0.21 &\\ 
\end{tabular}
\end{ruledtabular}
\end{table}

We first compare the computed effective masses with those from experiments \cite{Volm1996,Son2000}, as detailed in Table \ref{table1}. The effective masses are computed through polynomial fitting of the band structure near band extrema, as shown in Figure \ref{bzband}. Since the valence bands are very anisotropic and non-parabolic, quartic polynomials were used for them and the effective masses were computed from the quadratic term coefficients. The conduction band edge is at M point, as shown in Figure \ref{bzband}b, and the valley is anisotropic in all three directions. The electron effective mass is almost not affected by band structure method ($GW$ and HSE) or spin-orbit coupling (SOC), possibly due to the low intra-valley degeneracy. The transverse electron effective mass $m_\perp^e$ in basal plane, which is the average of effective masses along M-K and M-$\Gamma$ directions, is around 0.39. The longitudinal electron effective mass $m_\parallel^e$ along $c$-axis (M-L direction) is calculated to be around 0.33. All components of electron effective masses are very close to the experimental values from cyclotron resonance measurements \cite{Volm1996}. On the contrary, the hole effective mass is strongly affected by the SOC, while $GW$ shows very small improvement over HSE. The measured transverse hole effective mass $m_\parallel^e$ of 0.66±0.02 very close to the predicted heavy hole ($hh$) effective mass from $GW$+SOC calculation. The measured longitudinal hole effective mass of 1.75±0.02 is 17\% heavier than the predicted value. Considering the low measurement temperature (4 K), only $hh$ should be occupied and contributing to the cyclotron resonance signals. Overall, the electron effective mass predictions are closer to the experimental counterparts than those for holes, similar to the case of Si \cite{Ponce2018}.


\begin{figure}[h]
\includegraphics[width=16.4cm]{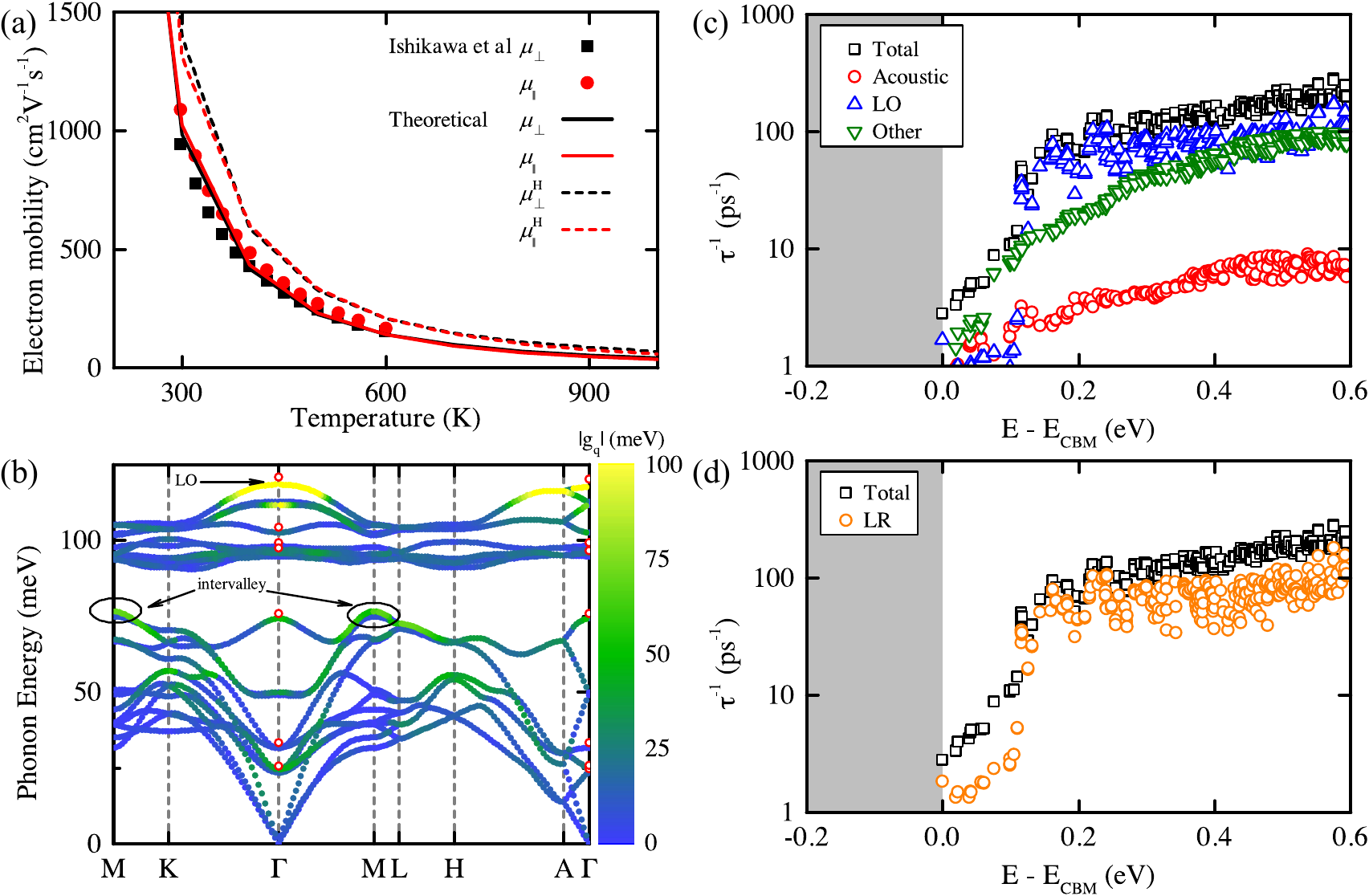}
\caption{\label{n_type} (a) The anisotropic electron mobilities compared with experimentally measured values by Ishikawa et al \cite{Ishikawa2021}. (b) The mode-resolved electron-phonon interactions $|g_{mn\nu}(\mathbf{k},\mathbf{q})|$ with the initial state $m\mathbf{k}$ at conduction band edge (M point) and $\mathbf{q}$ along high-symmetry path. The red circles mark the phonon energy measured at $10 ^\circ C$ \cite{Bauer2009}. Contributions of acoustic phonons, longitudinal optical (LO) phonons, and other optical phonons are compared with the total electron scattering rate at 300 K in (c). The long-ranged (LR) \textit{e-ph} interactions contributions are shown in (d).}
\end{figure}

\begin{figure}[h]
\includegraphics[width=16.4cm]{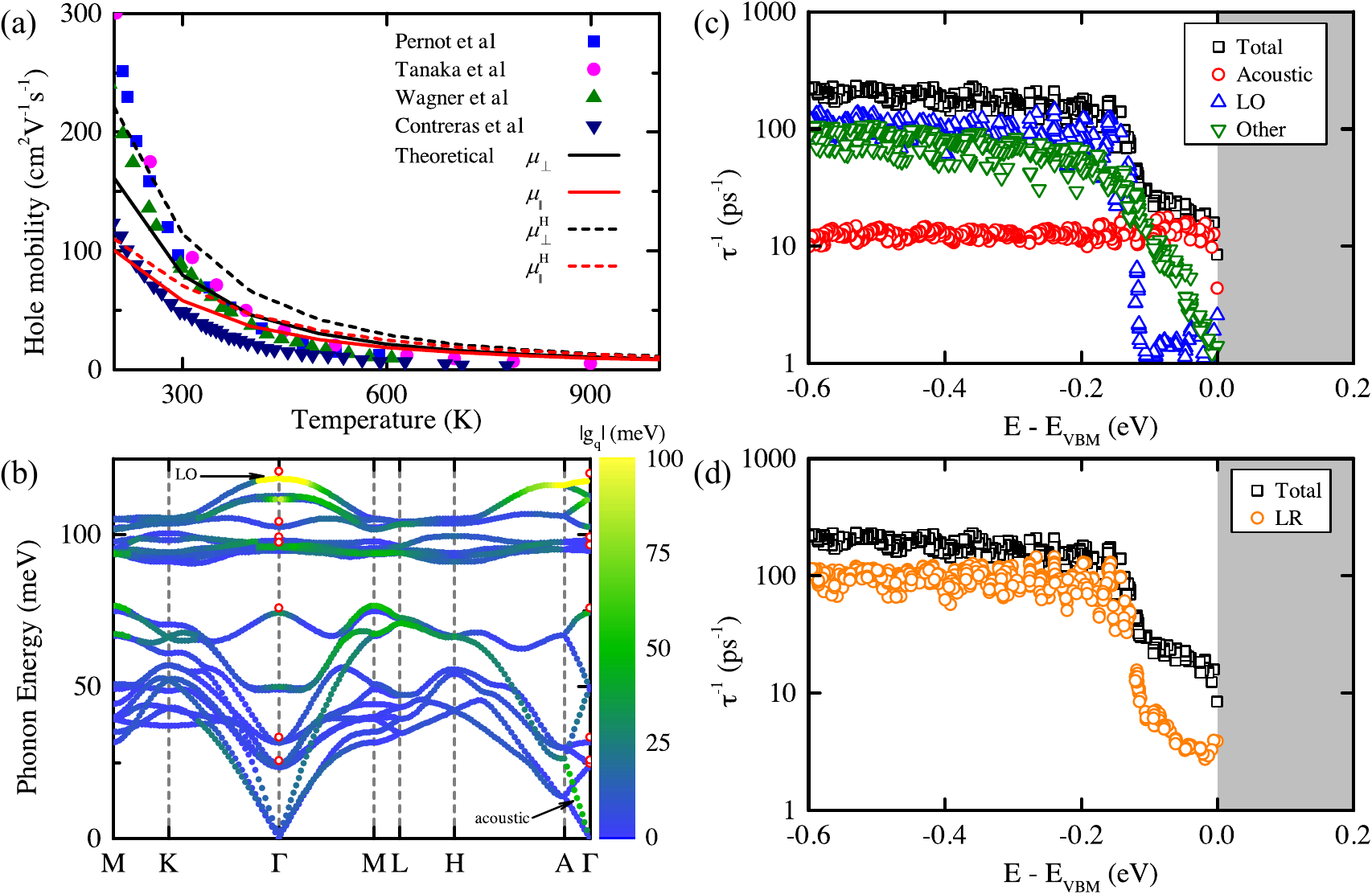}
\caption{\label{p_type} (a) The anisotropic hole mobilities compared with experimental measurement from literature\cite{Pernot2005,Tanaka2018,Wagner2002,Contreras2017}. (b) The mode-resolved electron-phonon interactions $|g_{mn\nu}(\mathbf{k},\mathbf{q})|$ with the heavy hole state at valence band edge ($\Gamma$ point) being the initial state $m\mathbf{k}$ and $\mathbf{q}$ along high-symmetry path. The red circles mark the phonon energy measured at $10 ^\circ C$ \cite{Bauer2009}. Contributions of acoustic phonons, longitudinal optical (LO) phonons, and other optical phonons are compared with the total hole scattering rate at 300 K in (c). The long-ranged (LR) \textit{e-ph} interactions contributions are shown in (d).}
\end{figure}

\subsection{Intrinsic mobility}

By solving the full-band BTE iteratively, we compute carrier mobilities using Eq. \eqref{eq:rh} to \eqref{eq:mu}. By increasing the density of $\mathbf{k/q}$ Brillouin zone mesh sampling from $15 \times 15 \times 5$ to $75 \times 75 \times 25$, we observe a rapid convergence of the drift mobilities. The Hall mobility, on the other hand, converges slower with a linear trend in $\mathbf{k}$-mesh density $1/N_k$ towards $N_k\to\infty$. Therefore, we linearly extrapolate the mobility data towards $N_k\to\infty$ following Ponc\'e et al \cite{Ponce2021} using the three points with densest $\mathbf{k}$-mesh, as detailed in Appendix.

We first compare the computed anisotropic electron mobilities with experimental measurement on an epitaxial sample fabricated on SiC (11$\overline{2}$0) substrate with a donor density of $2.1\times10^{15} \text{cm}^{-3}$ \cite{Ishikawa2021}. As shown in Figure \ref{n_type}(a), the experimentally measured Hall mobilities are slightly lower than the predicted phonon-limited Hall mobilities and close to the predicted drift mobilities. This is expected as additional scattering mechanisms such as impurities are not included in the calculations. The temperature dependence is also studied by fitting with a power-law relation, $\mu \propto T^{-\beta}$. The typical $\beta$ value for lightly n-doped 4H-SiC is 2.4 to 2.8\cite{Kimoto2014}, while Ishikawa et al \cite{Ishikawa2021} found an decreasing $\beta$ with increased doping concentration with highest $\beta = 2.67$ for $\mu_{\parallel}^{H}$ and $\beta = 2.58$ for $\mu_{\perp}^{H}$ at donor concentration of $2.1\times10^{15} \text{cm}^{-3}$. The calculated $\beta$ in this work is $2.82\pm0.03$ for $\mu_{\parallel}^{H}$ and $2.83\pm0.03$ for $\mu_{\perp}^{H}$ which is expected to be the intrinsic limit. The stronger $T$-dependence of intrinsic mobility is expected as defect-scatterings are absent and the strongly $T$-dependent phonon-scattering is the only mobility-limiting factor here.

Next we investigate the mode-resolved contributions to the \textit{e-ph} scatterings. As shown in Figure \ref{n_type}(c), most of the scatterings for high-energy electrons are contributed by the optical phonons, particularly the polar longitudinal optical (LO) phonons. Near the conduction band edge, the LO phonon contribution reduces due to energy mismatch and the low-energy electrons are mainly scattered by other optical phonons. Moreover, by computing the scattering rate only with long-ranged (LR) interactions, i.e. those from dipolar and quadrupolar \textit{e-ph} interactions, LR contribution is revealed to be important for both high- and low-energy electrons as shown in Figure \ref{n_type}(d). For further clarification, the \textit{e-ph} interactions between conduction band edge state at M point and phonons along high-symmetry path are calculated and visualized in Figure \ref{n_type}(b). It is evident that the electron-LO-phonon interactions are the strongest due to the diverging Fr\"ohlich interaction from non-zero Born effective charges whose average values are $\overline{Z_{\parallel}^*}\approx 2.78$ and $\overline{Z_{\perp}^*}\approx 2.68$, respectively. Additionally, the optical phonons around 75 meV at M point, which could scatter the electrons at one M valley to another M valley, is also significant and becomes another major scattering mechanism apart from the Fr\"ohlich interaction. \textit{e-ph} interactions with other intervalley phonons branches are much weaker. The interactions with acoustic phonons are rather weak, which could be attributed to the \textit{floating} nature of conducting electrons in SiC, whose wave functions are away from atoms or bonds \cite{Matsushita2012}. The short-ranged part of electron-ion interactions can be reduced in this case resulting in weaker deformation potential.

The hole mobilities are also computed and compared with experimental values from several reference\cite{Pernot2005,Tanaka2018,Wagner2002,Contreras2017}, as shown in Figure \ref{p_type}(a). The experimental data were all measured for epilayers on (0001) or slightly off-axis (0001) substrate. Therefore, the measured mobility should be close to the in-basal-plane component and are compared with $\mu_{\perp}$. While the measured mobilities are close to or lower than the computed Hall mobility $\mu_{\perp}^{H}$ around room temperature, at higher temperature the measured mobility decreases at a faster rate than the predictions. Previous studies attributed the strong decrease partially to the reduced hole Hall factor at high temperature, which can be further reduced by high doping levels\cite{Asada2016,Tanaka2018}. The impact of \textit{e-ph} interactions on Hall factor will be further discussed in the next session.

The hole scattering rates are also decomposed into different phonon branches, as shown in Figure \ref{p_type}(b) to \ref{p_type}(d). The holes at valence band edge are mainly scattered by acoustic phonons, while at higher energy above 0.1 eV the optical phonon scatterings are dominant since the emission of high-frequency optical phonons become available in this region. By comparing the contributions from acoustic phonons and LR \textit{e-ph} interactions, it is evident that the short-ranged deformation potential is dominant while the long-ranged piezoelectric scattering from non-zero quadrupolar coupling with acoustic phonons is weak for holes in 4H-SiC at 300 K. The strong deformation potential scattering arises from the $sp^3$ bond nature of the valence electrons, which can strongly couple with the atomic displacement through bond distortion. Although the intermediate-frequency intervalley phonons still couples strongly with holes, there is no final states available for scattering as the valence band edge is centered at the $\Gamma$ point. Therefore, the holes are mainly scattered by intravalley acoustic and optical phonons.

\subsection{Hall factor}
Using the fully \textit{ab initio} band structure without resorting to approximate parabolic or $k\cdot p$ band models, we first compute the Hall factor tensors in the constant relaxation time approximation without considering the scattering mechanisms. In this case, we eliminates the impact of scattering and their deviation from $r_H=1$ solely reflects the realistic band structure, i.e. the band multiplicity, anisotropy, and non-parabolicity.

\begin{figure}
\includegraphics[width=8.6cm]{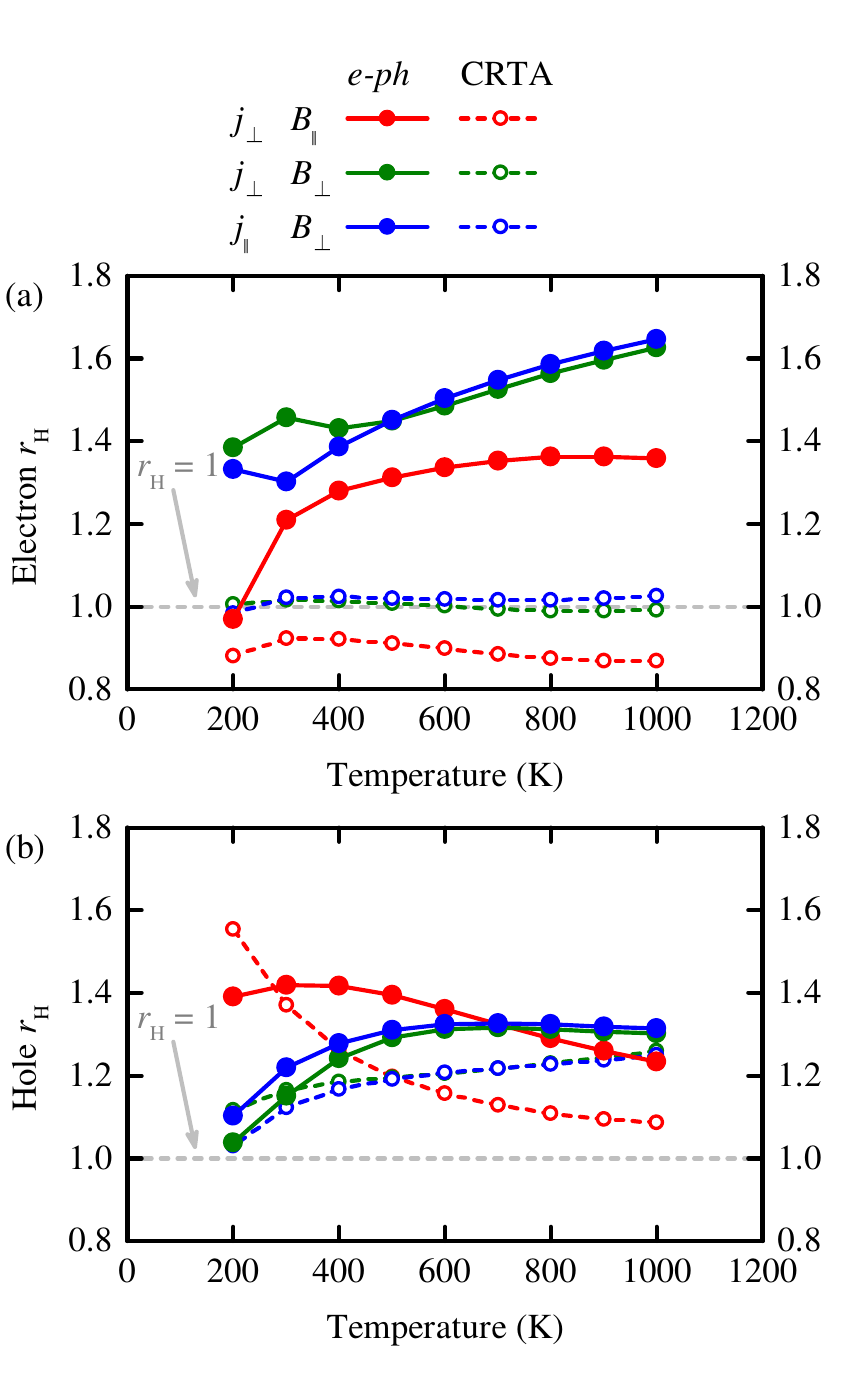}
\caption{\label{crta_hall} The Hall factor $r_{\text{H}}$ for (a) electrons and (b) holes, for current $j$ and magnetic field $\mathbf{B}$ parallel to ($\parallel$) or perpendicular to ($\perp$) the $c$-axis.}
\end{figure}

\begin{figure}
\includegraphics[width=8.6cm]{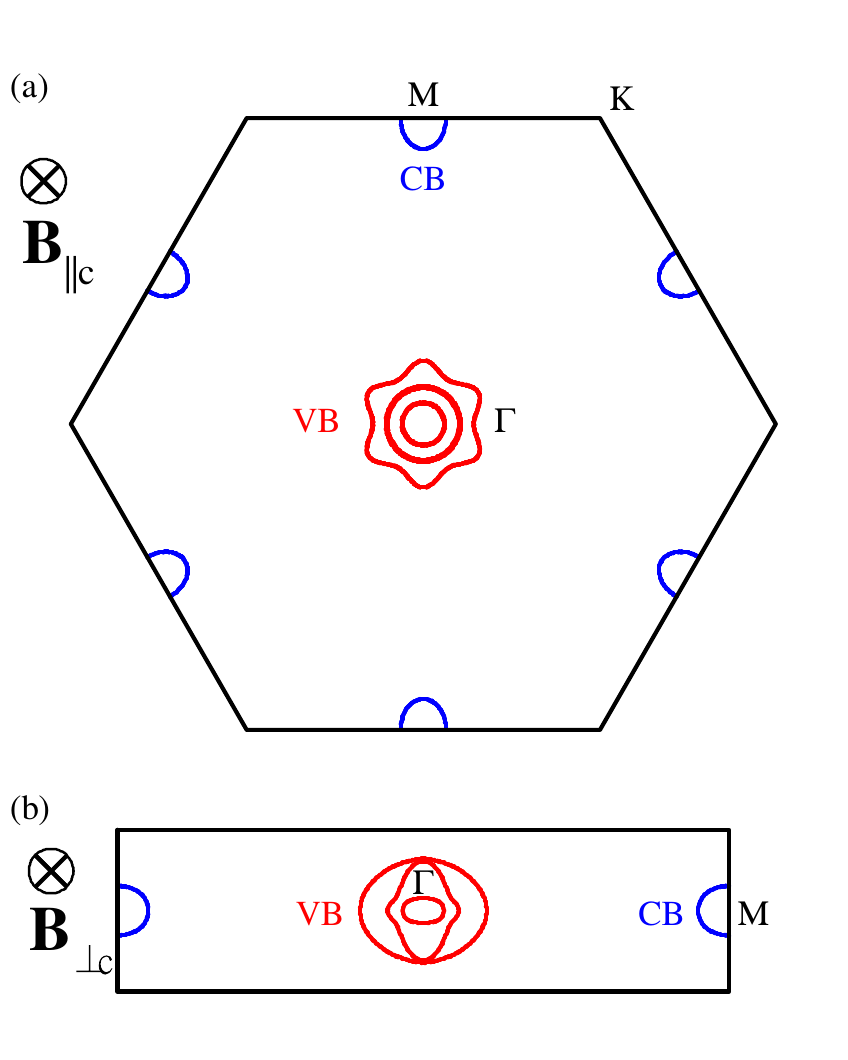}
\caption{\label{fs} The equienergy lines of conduction (blue) and valence (red) bands in a $\mathbf{k}$-plane perpendicular to the applied magnetic field, with (a) $\mathbf{B}$ along $c$-axis and (b) $\mathbf{B}$ perpendicular to $c$-axis, respectively. The equienergy lines are 0.1 eV above the conduction band edge and below the valence band edge.}
\end{figure}

As shown in Figure \ref{crta_hall}(a), the electron Hall factors are not significantly temperature-dependent. When the magnetic field $\mathbf{B}_{\parallel}$ is applied along $c$-axis, $r_{\text{H}}$ is around 0.9. The deviation from 1 is attributed to the in-plane anisotropy of the conduction valley, where the electron effective mass along M-$\Gamma$ direction is about twice that of M-K direction, leading to the elliptic equienergy line show in Figure \ref{fs}(a). When $\mathbf{B}_{\perp}$ is applied, the mass anisotropy in the planes perpendicular to $\mathbf{B}_{\perp}$ is reduced because $m_{\parallel}^{e}$ is between $m_{\text{M}\Gamma}^{e}$ and $m_{\text{M}\Gamma}^{e}$. Therefore, the Hall factors with $\mathbf{B}_{\perp}$ are very close to 1 as shown in Figure \ref{crta_hall}(a).

The case of holes is more complicated. The $hh$ and $lh$ bands are energetically close and are all involved in the transport process at the temperatures studied here. Moreover, these bands are very anisotropic and non-parabolic in the vicinity of valence band edge, as revealed by the effective masses and equienergy line shape (Figure \ref{fs}). Therefore, a deviation from unity is naturally expected even without any specific scattering mechanism as in CRTA. Indeed, the computed hole Hall factors are all away from unity. Interestingly, we observed different temperature-dependent Hall factors when magnetic fields are applied in different directions: when $B_{\parallel}$ is applied, $r_{\text{H}}$ decreases as temperature increases; when $B_{\perp}$ is applied, $r_{\text{H}}$ slightly increases at higher temperature. Such behaviours can be traced back to the equienergy line shapes of different bands. As illustrated in Figure \ref{fs}, in the (0001) plane, the low-energy heavy-hole $hh$ bands shows complex, non-elliptic equienergy line, while the high-energy $lh$ bands have almost isotropic, circular equienergy line. Therefore, as temperature increases, $lh$ bands participates into the transport process which explains the $r_{\text{H}}$ approaching 1 at high temperature with $B_{\parallel}$. In the case of $B_{\perp}$, as shown in Figure \ref{fs}(b), the $hh$ bands are elliptic while the $lh$ band becomes non-elliptic. Therefore, $r_{\text{H}}$ deviates further from 1 as temperature increases.

\begin{figure}
\includegraphics[width=8.6cm]{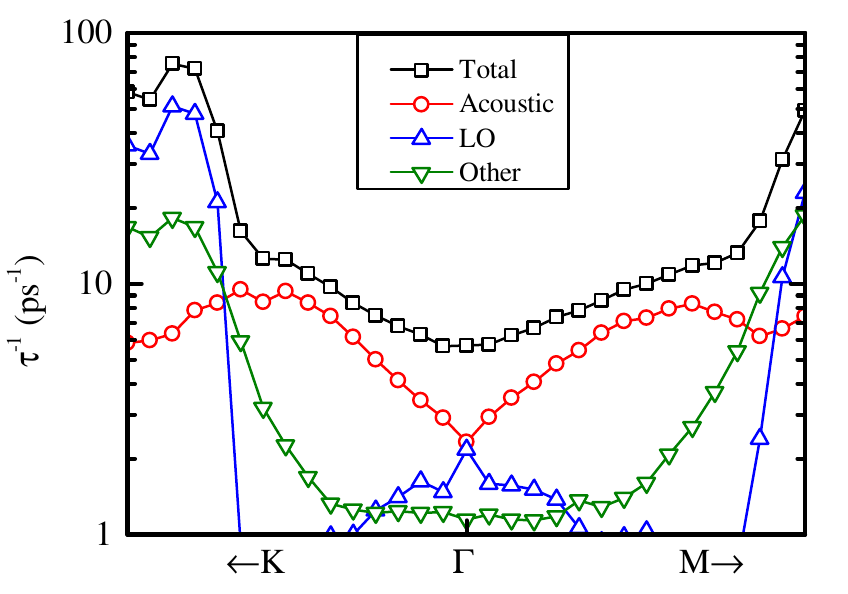}
\caption{\label{val_tau_path} The mode-resolved scattering rate of holes in heavy hole band near valence band edge. No significant anisotropy is observed for either the total scattering rate or contributions from individual phonon modes to the band edge holes.}
\end{figure}

When the \textit{e-ph} interactions are included and the BTE is self-consistently solved, the Hall factor $r_{\text{H}}$ changes drastically. The electron $r_{\text{H}}$ increases to as high as 1.6 at 1000 K and remains anisotropic depending on the direction of applied magnetic field. Specifically, when the magnetic field is applied perpendicular to $c$-axis, the Hall factor is almost isotropic. But the value changes significantly when magnetic field is switched to $B\parallel c$-axis, from around 1.6 to around 1.4 at high temperature. Additionally, the electron $r_{\text{H}}$ is increasing with temperature in contrast to the CRTA results which are almost constant across the studied temperature range. Similarly, the hole Hall factor is also different from CRTA results, but the change is less significant. This is attributed to the energy-dependent $\tau$, whose contribution can be approximately described in isotropic case by the Hall scattering factor $r_s=\frac{\langle\langle\tau^2\rangle\rangle}{\langle\langle\tau\rangle\rangle^2}$ where $\langle\langle \tau^i\rangle\rangle=\sum_{nk}\varepsilon_{nk}f_{nk}\tau_{nk}^i/\sum_{nk}\varepsilon_{nk}f_{nk}$\cite{Lundstrom2000}. The computed $r_s$ is 1.22 for electrons and 1.09 for holes at 300 K, which also reflects the different distributions of $\tau$ for electrons and holes near the band edge. The difference arises from both the difference in major scattering mechanisms, as discussed in previous section, and the difference in band structure complexity. Due to the strong anisotropy, non-parabolicity and band multiplicity, the available scattering phase space is more complex than the simple parabolic conduction band. This not only leads to the difference in CRTA $r_{\text{H}}$, but also different $r_s$ in the presence of \textit{e-ph} interactions.

We note that in previous literatures, the Hall factors (deduced from experimental Hall measurement, donor/acceptor density and ionization energy) are lower than the BTE-predicted values in this work. Considering the agreement between theoretical and experimental effective masses, the discrepancy likely arises from the factors other than the band structure, including but not limited to scattering. For example, Asada et al \cite{Asada2016} revealed experimentally that the Hall factors showed significant reduction with increased Al-doping in 4H-SiC, suggesting that doping can be a factor that lowers the Hall factor. Tanaka et al \cite{Tanaka2018} computed the p-type 4H-SiC Hall factor and mobility using BTE in relaxation time approximation with simplified phonon and impurity scattering model in conjunction with adjustable parameters. They assumed that the non-polar optical phonon scattering may lead to highly anisotropic scattering rate in the Basal plane direction ($\perp$) and reduces the Hall factor. However by computing the phonon mode-resolved contributions to the total scattering rate, as shown in Figure \ref{val_tau_path}, it can be seen that both the total and mode-resolved scattering rates are almost isotropic in the basal plane. This suggests that the transport anisotropy is likely dominated by the band structure. Therefore, experimentally observed small Hall factors may not be explained by anisotropic phonon-limited scattering rate. Alternative explanations could be impurity scatterings, dislocation scatterings, uncertainty in the estimated carrier concentrations in experimental works using hydrogenic model, among others.

\section{Conclusions}
In summary, we studied the phonon-limited electron and hole transport behaviors of the 4H silicon carbide by solving the Boltzmann transport equation in conjunction with Wannier interpolation of band structure and electron-phonon interactions from density functional (perturbation) theory calculations. The resulting effective masses and carrier mobilities agree well with experimental measurement. The anisotropy of band structure, electron-phonon scattering, and carrier mobilities are investigated in details. Spin-orbit interactions must be included to correctly reproduce the experimental valence band structure, while both HSE and GW calculations could excellently reproduce the experimental effective masses. We showed that the anisotropy of electron effective masses and mobilities is weaker than that of the holes, while both the electron and hole Hall factors are strongly direction- and temperature-dependent. The Hall factors significantly deviate from 1, which is explained by the energy-dependent electron-phonon scattering rate and anisotropic band structure. The results clarified the role of electron-phonon interactions in the transport phenomena of the technologically-relevant 4H-SiC.

\begin{acknowledgements}
This work was supported by the "Pioneer" and "Leading Goose" R\&D Program of Zhejiang (Grant No. 2022C01021). We are grateful for the computational resources provided by the National SuperComputer Center in Tianjin and the ZJUICI Supercomputer Platform.
\end{acknowledgements}

\appendix
\section{Wannier interpolation of band structure}

To test the Wannier interpolation reliability, we computed the band structure along high-symmetry path using both direct DFT calculation and Wannier interpolation from a $8\times 8\times 2$ coarse $\mathbf{k}$-mesh. The agreement is excellent for both conduction and valence bands. Therefore, the same Brillouin zone sampling is consistently used throughout this work. The Wannier functions are $p$-like orbitals centered on all C atoms for valence bands, and $sp^3$-like orbitals centered on all Si atoms for conduction bands, respectively.

\begin{figure}
\includegraphics[width=8.6cm]{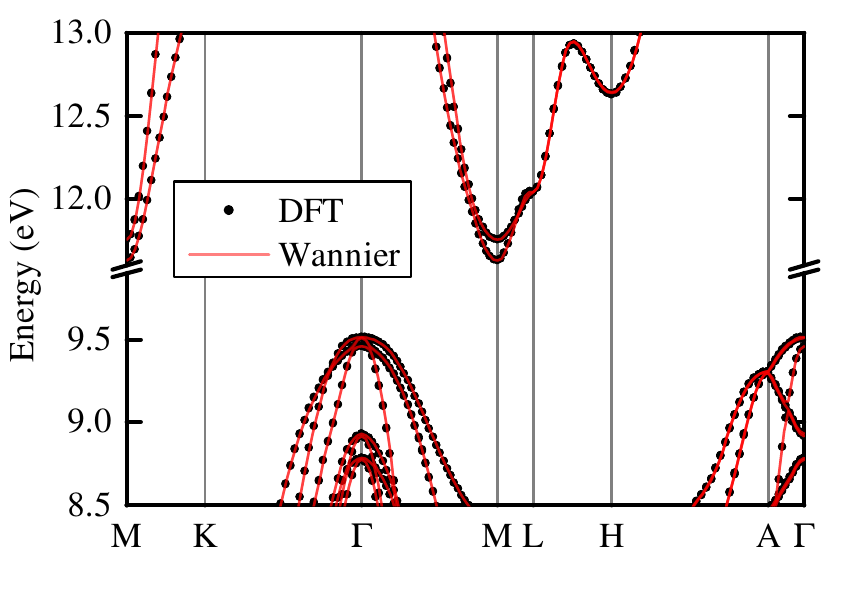}
\caption{\label{FR_band_wan} Comparison between band structures computed using direct DFT calculation and using Wannier interpolation starting from a $8\times 8\times 2$ $\mathbf{k}$-mesh.}
\end{figure}

\section{Mobility convergence with respect to the Brillouin zone sampling}

\begin{figure}
\includegraphics[width=8.6cm]{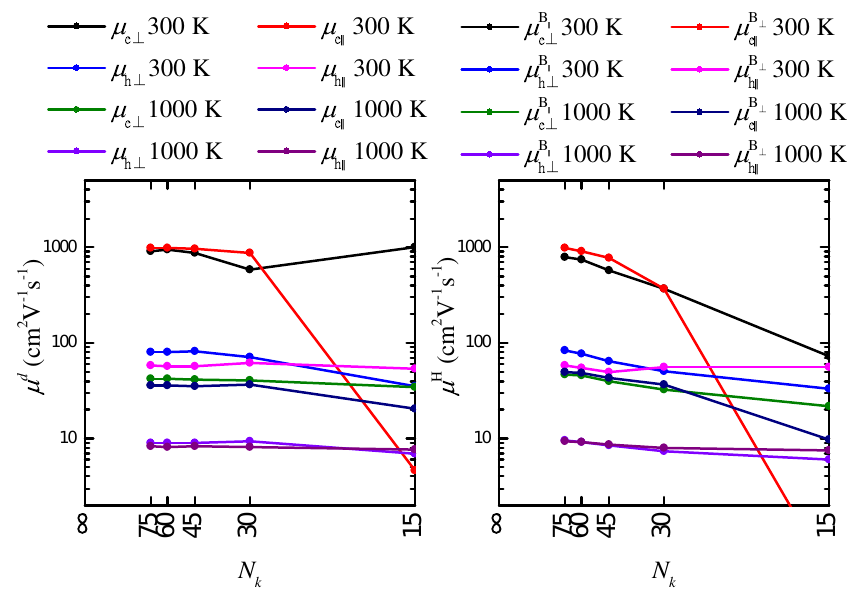}
\caption{\label{conv_mu} Convergence of drift mobility $\mu^d$ and Hall mobility $\mu^H$ with respect to Brillouin zone sampling point number along one axis $N_k$. The same grid was used for both electron $\mathbf{k}$ and phonon $\mathbf{q}$.}
\end{figure}

Ideally, the mobility from BTE could converge to a constant value as the Brillouin zone sampling becomes infinitely dense. However, this is computationally unattainable. Therefore, a finite Brillouin zone sampling grid must be used and its convergence should be tested. As shown by Ponc\'e et al, in many cubic semiconductors both the drift and Hall mobility approaches the converged value linearly in $1/N_k$ with $N_k$ being the fine $\mathbf{k/q}$ sample size along on direction. Here we also observed similar phenomena in the hexagonal 4H-SiC. As shown in Figure \ref{conv_mu}, both the drift mobility $\mu^d$ and Hall mobility $\mu^H$ steadily approaches their respective converged values. Therefore, we used the densest three sampling grids for the linear extrapolation towards $N_k \to \infty$.

\bibliography{hall}

\end{document}